# Enhancement of Raman Light Scattering in Dye-Labeled Rat Glioma Cells by Langmuir-Blodgett CNT-Bundles Arranged on Metal-Containing Conducting Polymer Film

A.S. Egorov[1], V.P. Egorova[2], H.V. Grushevskaya[3*], V.I. Krot[3], N.G. Krylova[3], I.V. Lipnevich[3], T.I. Orekhovskaya[4], and B.G. Shulitsky[4]

*Abstract*—We have fabricated layered nanocomposite consisting of a nanoporous anodic alumina sublayer (AOA), an ultrathin metal-containing polymer Langmuir-Blodgett (LB) film coating AOA, and multi-walled carbon nanotube (MCNT) – bundles which are arranged on the LB-film. MCNTs were preliminarily chemically modified by carboxyl groups and functionalized by stearic acid. We have experimentally observed an enhancement of Raman light scattering on surface plasmons in the LB-monolayers. This enhancement is due to charge and energy transfer. We demonstrate that propidium iodide (PI) fluorescence is quenched by the MCNT-bundles. A method of two-dimensional system imaging based on the MCNT-enhanced Raman spectroscopy has been proposed. This method has been applied to visualize focal adhesion sites on membranes of living PI-labeled rat glioma cells.

## I. Introduction

It is known [1], [2] that a living cell has subcellular structures such as focal contacts, that provide cohesive bonds with hydrophobic sites of other cells or extracellular matrix. Ends of focal contacts are a set of integrin receptor zones. The hydrophobic regions of biological structures are investigated by means of non-specific fluorescent probes, such as propidium iodide (PI) used for DNA-containing structures. Such non-specific probes for the focal contacts are not known. Meanwhile, the ordinary method of fluorescent-labeled specific antibody needs a fixation of cells and a permeabilization of cellular membrane that distort subcellular structures.

A.S. Egorov[1] is with Institute of Physiology, Nat. Acad. Sci. of Belarus, Belarus (e-mail: biblio@fizio.bas-net.by).
V.P. Egorova[2] is with Belarusian State Pedagogical University, Belarus (e-mail: valentinaegorova@yahoo.com).
H.V. Grushevskaya[3], V.I. Krot[3], N.G. Krylova[3], I.V. Lipnevich[3] are with Belarusian State University, Belarus (corresponding author's phone: +375172095122; e-mail: grushevskaja@bsu.by, nina-kr@tut.by).
T.I. Orekhovskaya[4], B.G. Shulitsky[4] are with Belarusian State University of Informatics and Radioelectronics, Belarus (e-mail: taisa-o@tut.by, shulitski@bsuir.by).

Surface-enhanced Raman spectroscopy (SERS) is a promising method with improved sensitivity which allows to detect structural and functional properties of nanoscale structures in a data-rich manner. SERS usage allows to obtain Raman spectra of such nanostructures as LB-films deposited on a surface with silver islands [3]. In papers [4], [5], the surface-enhanced Raman spectroscopy has been proposed to study cells and subcellular structures on a monolayer of gold or silver nanoparticles coated by surfactant. However, the usage of metal surface limits SERS applicability in biotechnology [6]. The surface-modified gold or silver nanoparticles are accumulated inside cells or selectively bind with membrane receptors, disrupting cell functioning. Another problem of the method is to reproducibly prepare high-sensitive SERS-active nanostructures with narrow distribution in enhancement factor values [7], [8]. Size of gold or silver nanoparticles (more than 20 nm) is large comparing with a size of focal contact integrin receptor (about 10 nm) [9]. Height of receptor external part is about 10–15 nm [1]. Because of this, the resolution of known methods which utilize metal nanoparticles is not sufficient to visualize distribution of cellular supporting structures (focal adhesion sites). This explains a large spread of experimental data on focal adhesion site sizes: the width is about 1–3 μm and the length is about 2–10 μm [1], [2].

The use of carbon nanotubes (CNT) for SERS [10] is hampered due to the lack of free charge carriers at the Fermi level in graphene and graphene-like materials [11]–[14]. However, a charge and energy transfer from dye-labeled DNA to CNT leads to quenching of DNA scattering, and, respectively, to an appearance of transport of free charge carriers in CNTs [15]–[17]. The resonance of CNT plasmon at eigenfrequencies of silicon surface vibrations gives rise to a phenomenon of CNTs-enhanced Raman scattering in Si [15]. In this paper we shall demonstrate similar charge and energy transfer from LB-bundle of carbon nanotubes with a small number of walls (MCNTs) to ultrathin conducting metal-containing LB-film [18] which is deposited on a surface of nanoporous anodic alumina. In this paper we shall demonstrate also that the propidium iodide fluorescence quenching is observed on LB-MCNT-bundles deposited on AOA with the LB-film.

The goal of the paper is to reveal and explain a phenomenon of MCNT-enhanced resonance Raman light scattering and to apply this phenomenon for visualization of cell – matrix adhesion sites of living rat glioma cells stained with propidium iodide. The paper is organized as follows. In Section II materials and methods of investigation are described. In Section III.A we perform a structural analysis of proposed layered nanoheterostructures consisting of a AOA sublayer, an ultrathin metal-containing polymer LB-film coating AOA, and MCNT-bundles which are arranged on the LB-film. In Section III.B we establish the appearance of free charge carriers in MCNTs when they are complexified with double-stranded (ds) DNA and PI molecules. In Section III.C we represent a method of two-dimensional (2d) system imaging. This method is based on the enhancement of Raman light scattering in the metal-containing conducting LB-film by LB-MCNT-bundles deposited on AOA. And at last the proposed method is used to visualize focal adhesion sites on membranes of cells attached to the surface of LB-MCNT-bundle. In Conclusion we summarize our findings.

## II. EXPERIMENTAL

### A. Materials

RNA and proteins fractions in high-purity double-stranded DNA (1.03 mg/ml in $10^{-5}$ M $Na_2CO_3$ buffer medium (TE-buffer)) were less that 0.1% (optical density ratio $D_{160}/D_{230}$ = 2.378 and $D_{160}/D_{280}$ = 1.866, respectively). MCNTs with diameters ranging from 2.0 to 10 nm and length of ~ 2.5 μm were obtained by the method of chemical vapor deposition (CVD-method) [19]. MCNTs were covalently modified by carboxyl groups and non-covalently functionalized by stearic acid molecules [20]. Complexes ds-DNA/MCNT were formed by ultrasonic treatment of alcoholic solution of mixture ds-DNA – MCNTs [21]. The complexes ds-DNA/MCNT were mixed with a solution of stearic acid in hexane. The resulting mixtures were homogenized by ultrasonic treatment to form hydrophobic (reverse) micelles of stearic acid with ds-DNA/MCNT complexes inside them [22].

A conducting oligomer 3-hexadecyl-2,5-di (thiophen-2-yl)-1H-pyrrole with chemically bounded hydrophobic 16-link hydrocarbon chain (DTP, thiophene-pyrrole) [18], [23] is utilized to synthesize coordination complexes of thiophene-pyrrole with iron atoms. The coordination complexes obtained by means of the LB-technique are the nanocyclic organomet compound of thiophene-pyrrole ligands with bivalent iron $Fe^{2+}$ in high-spin state [18], [23].

C6 rat glioma cells obtained from culture collection of Institute of Epidemiology and Microbiology (Minsk, Belarus) were grown on aligned MCNT arrays and LB-DTP-films in a Dulbecco's modified Eagle medium (DMEM) (Sigma, USA) supplemented with 10 % fetal bovine serum and $1 \cdot 10^{-4}$ g/ml gentamycin at 37 °C in a humidified 5 % $CO_2$ atmosphere. Propidium iodide (Sigma, USA), luminescence of which is high-intensive in hydrophobic environment, for example, when it intercalates into DNA sequence, but is low-intensive in hydrophilic medium (because of quenching by water molecules), has been used as a fluorescent probe [24]. Cells were stained with propidium iodide for 5 min.

All other reagents were of analytical grade.

### B. Methods

Fabrication of AOA films (thickness of 200-300 nm) and membranes (thickness of 20.0 μm) were carried out by means of two-stage anodic treatment in 10 % sulfuric acid solution at voltage of 10 V and temperature of 2 ºC [25]. The diameter of AOA pore was equal to 10 nm.

Langmuir monolayers are fabricated on an automated home-built Langmuir trough with controlled deposition on a substrate [18]. Langmuir monolayers of conducting polymer were obtained by compressing of the thiophene-pyrrole oligomer on a air / aqueous subphase interface. Preliminary, the solution of thiophene-pyrrole in hexane was dripped with help of a micropipette on a liquid subphase surface. The monolayers after their compressing were lifted by Y-type transposition at that the substrate is immersed into the water before spreading the solution. LB-monolayers were deposited by vertical lift at a constant surface pressure of 25 mN·m$^{-1}$ (with downward speed 4 mm·min$^{-1}$). Aqueous solution of ferric nitrate $Fe(NO_3)$ with pH=1.65 adjusted by addition of hydrochloric acid HCl was used as a subphase for thiophene-pyrrole monolayer formation. All salt solutions have been prepared with deionized water with resistivity 18.2 MΩ·cm.

Five metal-containing thiophene-pyrrole LB-monolayers in the high-ordered solid state were deposited on AOA which was preliminary hydrophilized by pure DTP dropped on its surface.

LB-monolayers fabricated from carboxylated MCNTs were preliminary functionalized by stearic acid. The functionalization consisted in preparation of stearic acid hydrophobic micelles with the hydrophilic carboxylated MCNTs inside them by sonication of a mixture from stearic acid and MCNTs in hexane. Then the micelles were dripped on subphase surface in LB-trough. Deionized water was used as a subphase for MCNT-monolayer formation.

Scanning electron microscopy (SEM) images were taken on LEO 1455 VP (Carl Zeiss, Gernamy) JEM-100CX. The accelerating voltage was 20 kV. Signals of reflected and secondary electrons were detected simultaneously. In order to determine crystallographic characteristics of crystallites the diffraction appliance to the scanning electron microscope – HKL EBSD Premium System Channel 5 was used (Oxford Instruments, England). To obtain a reflected electron diffraction (RED) patterns, a sample was placed on a special table, having a tilt angle to the horizontal of about 70 degrees. Electron probe was directed to the point of interest on a sample surface: elastic scattering of the incident beam causes the electrons to be deviated from the point directly below the sample surface and the bump on crystal planes from all sides. In cases when a Bragg diffraction condition is satisfied for the planes formed by atoms of the crystal lattice, two conical beams of diffracted electrons are formed for each family of crystal planes. These cones of electrons can be visualized by placing on their way a phosphorescent screen and after it a highly-sensitive camera for observation (digital CCD camera). In places where the cone-shaped electron beams intersect with the phosphorescent screen, they appear as thin bands called Kikuchi bands. Each of these bands corresponds to a specific group of the crystal planes. The resulting RED pattern consists of many Kikuchi's bands. With the help of Flamenko software the position of each Kikuchi's band was automatically determined, comparison with theoretical data and the conclusion on corresponding crystalline phase were performed as well as a three-dimensional crystallographic orientation was calculated.

Microdiffraction patterns and transmission electron microscopic images were obtained by means of transmission electron microscope JEM-100CX (JEOL, Japan) (TEM) at accelerating voltage of 100 kV. After the objects were previously deposited on a copper grid with a formvar polymer coating or on thin AOA membranes, structural analysis has been performed. AOA membranes twere made very thin by etching.

Spectral studies in visible range were carried out using a confocal micro-Raman spectrometer Nanofinder HE («LOTIS-TII», Tokyo, Japan–Belarus) with laser excitation at wavelengths 473 or 532 nm and vertical spatial resolution 150 nm. The immersion confocal spectroscopy with immersion objective has been used to image living cells. The images represent themself a rectangular area formed by 1500 – 1800 points with spatial size step 750 nm.

### III. RESULTS AND DISCUSSION

#### A. Structural Analysis

*Structural Analysis of MCNT and Micelles.* TEM-image of original carboxylated MCNTs is shown in fig. 1a. MCNTs have an "open"-type ends. Stearic acid functionalized MCNTs, rolled into a ball which one can see in fig. 1b, are inside micelles showed in fig. 1c. The hydrophobic micelles aggregate on the AOA surface (fig. 1e), as fig. 1d demonstrates.

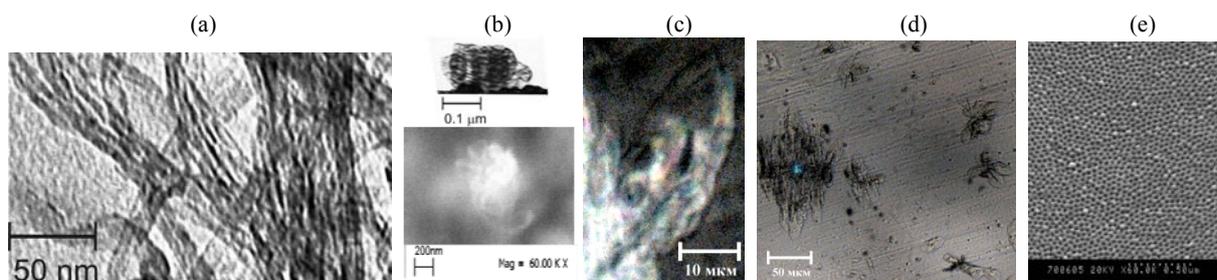

Fig. 1 TEM-image of non-functionalized carboxylated MCNT bundle (a) on formvar polymer coating; SEM- (b, down) and TEM- (b, up) images of the MCNTs in the interior of stearic acid micelle on etched AOA membrane and on copper grid edge, respectively; confocal microscope images of MCNT-containing stearic acid micelle (c) and micelle aggregates (d) on AOA membrane; SEM-image of AOA (e)

*Structural Analysis of LB-MCNT-Bundles.* A TEM-image of the metal-containing thiophene-pyrrole LB-film is shown in fig. 2a. Two monolayers of functionalized MCNTs were deposited on AOA, the surface of which was hydrophobized by the thiophene-pyrrole LB-film. In the fabricated nanocomposite the LB-DTP-film plays the role of a support for LB-MCNT-bundles. The morphology of LB-MCNT-monolayers is shown in fig. 2b. As one can see in fig. 2b, the high-ordered LB-MCNT-bundle is 2d horizontally-aligned MCNT array, ends of which are vertically anchored in the surface of conducting thiophene-pyrrole LB-film. The diffraction pattern of LB-MCNT-bundles on the surface of LB-DTP-film is represented in fig. 2c. Diffraction of the LB-MCNT-bundles also is characterized by Kikuchi bands [21]. This means that the LB-MCNT-bundles deposited on thiophene-pyrrole LB-film are crystallites. Further we will show that in the sites of contacts of vertical-aligned ends of MCNT-bundles with metal-containing LB-film, charge and energy transfer to LB-MCNT-bundles can occur.

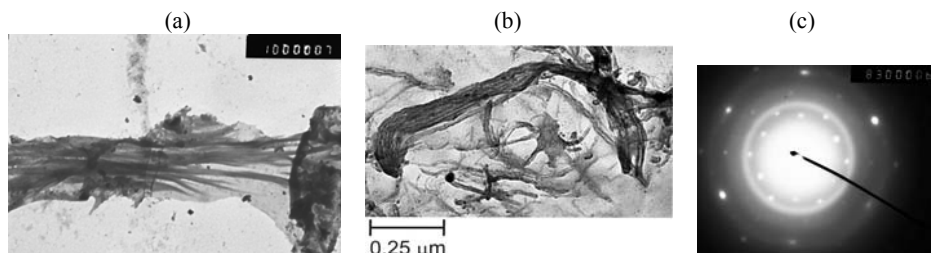

Fig. 2 TEM-image of metal-containing thiophene-pyrrole LB-film (a) on formvar polymer coating. TEM-image (b) and microdiffraction pattern (c) of LB-MCNT-bundle deposited on the metal-containing thiophene-pyrrole LB-film coating formvar-polymer support

*Structural Analysis of DNA/MCNT Complexes.* DNA/MCNT complex is a layer of electron-dense nucleotide self-organized on MCNT surface [26]–[28]. As shown in fig. 3, practically black nucleotide shell links ("glues") the dazzling white carbon nanotubes one to another, that creates conditions for charge transport and energy transfer. The sites in which ds-DNA glues one MCNT to another two MCNTs are pointed out by arrows in fig. 3. "Glued" each other MCNTs form macroscopic conducting paths.

### B. Raman Spectroscopy

*Raman Spectroscopy of MCNTs.* CNTs are a graphene-like material. CNTs, similarly to graphene, possess high electroconductivity [11]. The charge carriers in graphene are negatively and positively charged quasi-particle excitations. Electrons and holes in undoped (pure) graphene are localized in separate regions (so called "puddles") of cone valence zone [11] and do not appear in a Dirac point $K$ – the top of the Dirac cone. Although pure graphene has free charge carriers, its Fermi level lies in the point $K$ where charge carriers are absent because of graphene bipolarity.

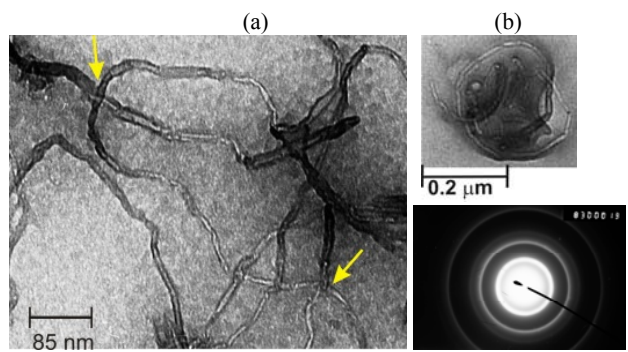

Fig. 3 TEM-images of ds-DNA/MCNT complexes in TE-buffer (a) and in the interior of stearic acid micelle (b, up); and microdiffraction pattern of the complexes (b, down) on formvar coating

Low-intensive Raman spectrum of dried original MCNTs on a clean Si support is shown in fig. 4a. Raman spectra of graphene and graphen-like materials have been studied in detail (see [29] and its references). The spectral bands $D$ and $D'$ in fig. 4 reveal the existence of defects in graphene lattice and correspond to optical transverse and longitudinal in-plane vibrations in the vicinity of point $K$ of the Brillouin zone. These phonons are nucleus oscillations in the field (term) of $\pi(p_z)$-electrons in valence band or $\pi^*(d)$-electrons in conduction band. The peak $D''$ is a longitudinal acoustic mode in the vicinity of point $K$. The spectral band $G$ originates also from in-plane carbon atoms vibrations, but in the electronic-vibrational term of $sp^2$–hybridized electrons. This resonance corresponds to optical surface phonons in the vicinity of point $\Gamma$ of the Brillouin zone. $D^3$ and $D^4$ are transverse and longitudinal acoustic branches of in-plane vibrations in the vicinity of point $\Gamma$. $2D$ is a peak of two-phonon absorption (doubled $D$ mode). Radial Breathing Mode (RBM) [30]–[31] for MCNT is observed in frequency range from 60 to 430 cm$^{-1}$ as shown in fig. 4.

It follows from comparing of fig. 4a with fig. 4b, Raman spectral lines $G$, $D$, $D'$, $D''$, and $2D$ for MCNTs being into the hydrophobic micelles and LB-MCNT-bundles are intensive due to their close packing. The comparison of Raman spectra of micellar MCNTs (curve 1 in fig. 4b) and original non-functionalized MCNTs (fig. 4a) demonstrates that the ratio $I_D/I_G$ of $D$ and $G$ peaks intensities $I_D$, $I_G$, respectively, are larger for micellar MCNTs. This rise in the defect peak intensity $I_D$ is due to the additional distortion of MCNT walls. LB-technique selects defectless MCNTs, as the ratio of $D$ and $G$ peaks intensities $I_D/I_G$ sharp falls in fig. 4b.

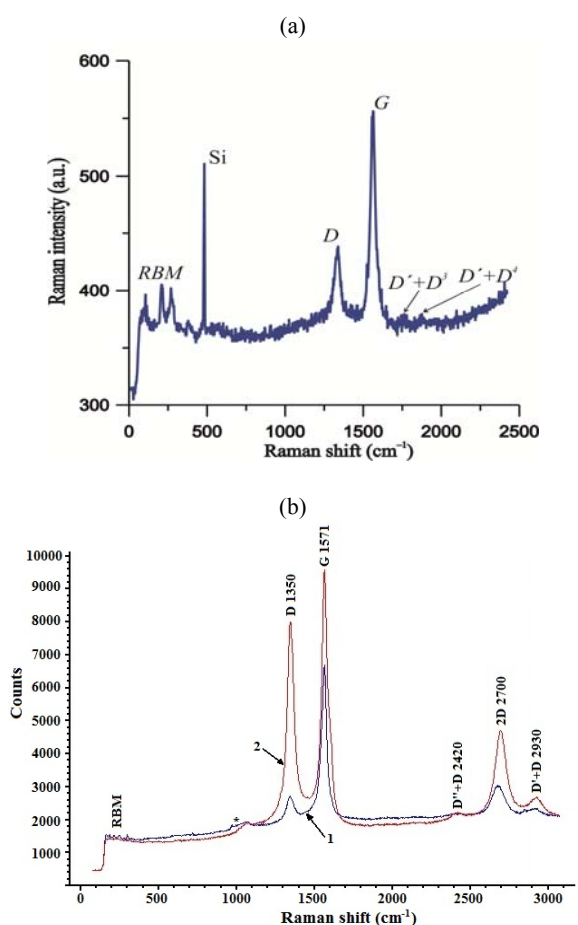

Fig. 4 Raman spectra (a) of original carboxylated MCNTs on Si, (b) of MCNT/stearic acid micelles (red curve 2 with high peak $D$) and of LB-MCNT-bundles (blue curve 1 with low peak $D$) on AOA. Excitation by green laser (532 nm) with power 20 (a) and 14.4 (b) mW. Symbols «Si» and «*» designate vibrational mode of Si and the laser peak of plasma scattering, respectively. (In color)

*Raman Spectroscopy of ds-DNA.* Raman spectrum of ds-DNA-containing micelles, which were dripped on a polished silicon support, has all characteristic lines of Raman light scattering in DNA, as shown in fig. 5a.

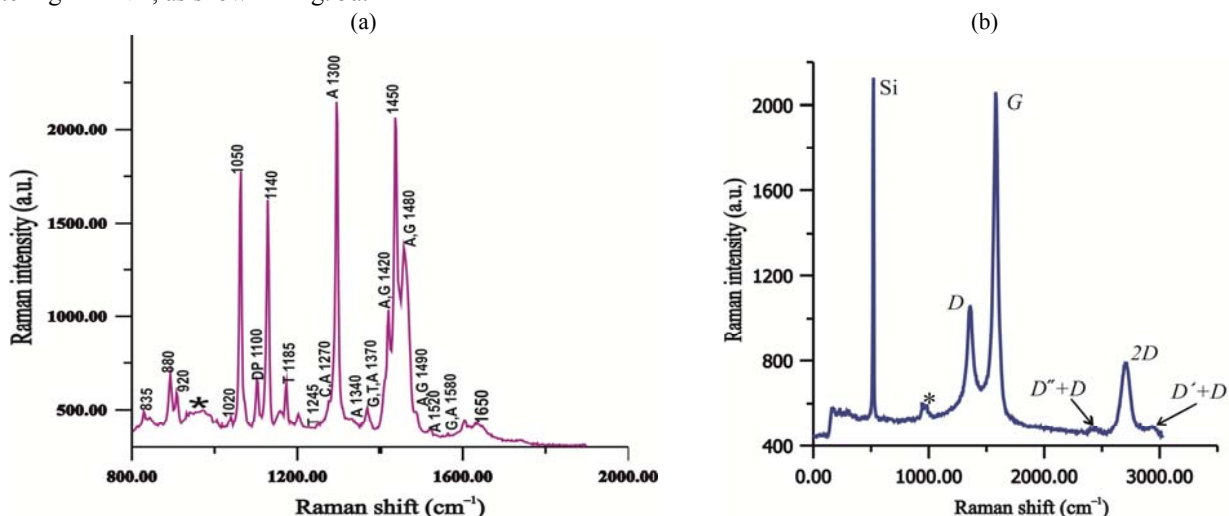

Fig. 5 (a) Raman spectrum of micelles formed in hexane solution of mixture from stearic acid with ds-DNA and dripped on Si. The numbers indicate the characteristic frequencies of DNA vibrations; adenine, guanine, thymine and cytosine are designated by A, G, T and C, respectively; DP denotes phosphodiester bond. Laser power was 20 mW, excitation wavelength was 532 nm. (b) Raman spectrum of hydrophobic stearic acid micelles with DNA/MCNT complexes on Si surface: excitation by blue laser (473 nm) with power 5.76 mW. Symbols «Si» and «*» designate the vibrational mode of Si and the laser peak of plasma scattering in silicon, respectively

*Surface-Enhanced Raman Scattering in DNA/MCNT Complexes.* Fig. 5b demonstrates Raman spectrum for a monolayer of the stearic acid micelles with ds-DNA/MCNT complexes inside. The dense packing of MCNTs in ds-DNA/MCNT complexes significantly intensifies the MCNT modes: $D$, $G$, $D'' + D$, $2D$, $D' + D$. The characteristic modes of ds-DNA are not observed in fig. 5b, and, consequently, the quenching of DNA Raman modes occurs. This is due to transfer of electric charge carriers and quasi-particle excitation energy from nucleotide bases to MCNTs. The probability of electron transfer from valence $\pi(p_z)$- orbital of nucleotides to $\pi^*(d)$-orbital of MCNTs is so high that Raman light scattering in DNA molecules is quenched completely as comparison of figs. 5a and 5b shows. MCNTs are effective quenchers of FAM-labeled DNA fluorescence (see [15] and its refs). Further we will demonstrate that MCNTs quench PI-fluorescence as well.

*MCNT-Enhanced Raman Scattering in Nanocomposite LB-MCNT-Bundles / LB-DTP-Film on AOA.* Raman spectra of metal-containing DTP-monolyers are shown in figs. 6 and 7. The most intensive line in Raman spectrum of Fe-containing thiophene-pyrrole compound is a pyrrole ring vibrational mode at frequency 1462 cm$^{-1}$ for non-annealed samples and at frequency 1472 cm$^{-1}$ for samples annealed at 100 °C. Besides peaks, in spectrum of metal-containing DTP-monolyers one can observe a wide plasmon band from 1600 to 2300 cm$^{-1}$ (fig. 7, curve 2) which is absent for AOA at exciting light intensities $I$ up to 3.0 mW [16]. It follows from comparing of the curves 1 and 2 in fig. 6, MCNTs enhance the intensities of the pyrrole peak and plasmon band (1600-2300 cm$^{-1}$). This rise is caused by two reasons. Firstly, defectless conducting areas of LB-DTP-films which are connected each other by MCNTs, form additional macroscopic conducting paths as well as in the case of complexes DNA/MCNT. But, unlike DNA/MCNT, the charge and energy transfer occurs from MCNT to LB-DTP-monolayers, and Raman modes of MCNTs are quenched.

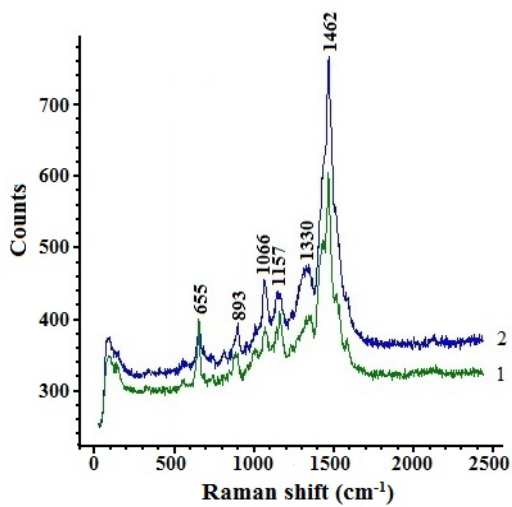

Fig. 6 Raman spectra of LB-films from five monolayers of DTP (1) coated by two monolayers of MCNT with stearic acid (2). Excitation was performed by green laser (532 nm) with power 1.2 mW

Secondly, Fe-containing LB-DTP-monolayer is a graphene-like bipolar material with voids (bubbles), size of which is about 8Å [18], [32]. Fe$^{2+}$ ions are placed in these bubbles. An additional electron density and energy transferred from MCNTs to Fe-containing LB-DTP-monolayers provide the

Fermi level shift in conductivity band. Owing to coincidence of "Dirac" Brillouin zone for electrons and holes in the Fe-containing LB-DTP-monolayer, some part of additional electron density annihilates and the released energy is spent on transition of charge carriers (negatively charged excitons) in the thiophene-pyrrole film conducting band. It leads to an increase of plasmon band intensity in frequency range 1600-2300 cm$^{-1}$.

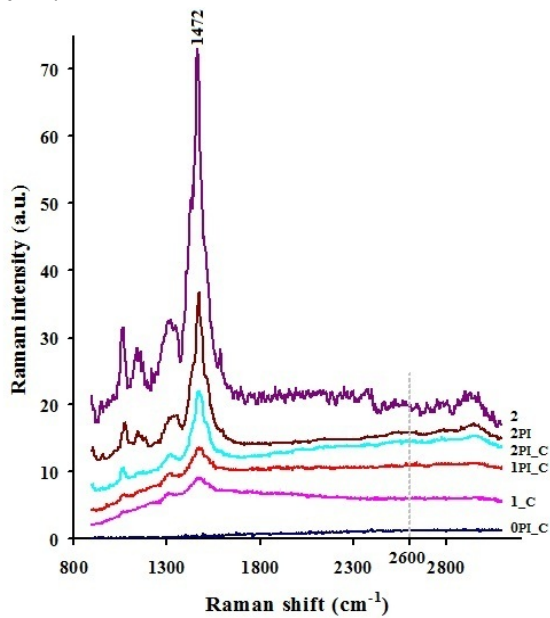

Fig. 7 Raman spectra of propidium iodide (curve 0PI_C), of five monolayers LB-DTP-films (curves 1_C, 1PI_C) coated by two monolayers of MCNTs with stearic acid (curves 2, 2PI, 2PI_C). The spectra represent themself sums of spectra over all points of scanning areas. "PI" denotes the presence of 1μM propidium iodide in the cuvette, "C" denotes the presence of cellular monolayer on the LB-film. Excitation was performed by green laser (532 nm) with power 0.06 mW. (In color)

At low pumping intensities the quantity of charge transferred from MCNTs to LB-DTP-monolayers is slightly. Owing to this smallness all additional electron density annihilates. At increase of laser excitation $H_{ext}$ the electron density transferred from MCNTs annihilates incompletely, and some part of it becomes an impurity. Since additional electrons occupy π*(d)-orbitals of atoms C, one can assume the following. These electrons are localized in the vicinity of high-spin $Fe^{2+}$-containing bubble, electron density of which has the holes on s- and d-orbitals. The resulting system is similar to the atom of large size – Rutherford atom. The multiphoton absorption through reflection diffraction on the bubble boundary leads to Hanle effect – mixing of π*(d)-orbitals with s-like orbitals from $Fe^{2+}$ electron shell with the next value of principal quantum number.

This mixed electron state of the Rutherford atom Fe is described by a wave function

$$|\Psi\rangle = |d\rangle + f(H_{ext})|s\rangle \qquad (1)$$

Let us consider the transition from p-like orbitals into the state (1) with total angular momentum $\vec{J} = \vec{j}_d + \vec{j}_s = \vec{L} + \vec{S}$; $J = 1$ at non-equilibrium spin $S = 0$ and orbital angular momentum $L = 1$. The Hanle effect leads to Zeeman splitting of the p-orbital. The transition from the state with momentum $J = 2(l–s) = 1$ with $l=1$ and $s=1/2$ is forbidden by angular momentum conservation law.

At large intensities of laser excitation, the oscillatory Hanle effect that leads to degeneration of p-like orbital is observed [32]. The electron shell of the Rutherford atom Fe consists of localized d-electrons with holes in the electron density. In this case, the Rutherford atom as a donor impurity can also add electron density in the vicinity of Dirac point $K$ via the resonant excitation of coherent negatively charged excitons $X^-$ in following way. Because p-electron transitions from degenerated p-like orbitals with $S = 0$, $L = 0$ into the mixed state $S = 0$, $L = 1$ are allowed, the excited p-electrons can be hybridized with π*(d)-orbitals of Rutherford atom to form "diffraction" negatively charged excitons $X^-$.

The emergence of hybridization gap leads to transition of d-electrons to π($p_z$)-orbitals of atoms C of thiophen–pyrrole ligands with emission of photons, energy of which is spent on the transition of charge carriers (negatively charged excitons $X^-$) from valence band to the conductivity one. As result, the quasi-particle excitations $X^-$ resonate with s-electrons of carbon atoms involved in $sp^2$-hybridization, and, respectively, exciton energy scattered in Raman bands of thiophene having donor atoms S is transferred in the chain of conjugated double bonds.

The d-electron levels of the donor impurity Rutherford atom Fe appear in the process of hybridization of the mixed state orbitals ($S = 0$, $L = 1$) with π*-orbitals of atoms C of thiophene–pyrrole ligands in conductivity band (near point $K$ of Brillouin zone) through the scattering of the mixed state on the vibrational modes of ligand pyrrole rings with acceptor atoms N.

Thus, the charge and energy transfer from MCNTs does not only prevent the charge transport in the LB-DTP-monolayes, but even increases the number of free charge carriers of the last.

*MCNT-Enhanced Resonance Raman Spectroscopy with Propidium Iodide.* Raman light scattering in fluorescent probes PI is resonant one. Propidium iodide fluoresces in a wide range of wavelengths from 585 to 685 nm at excitation by green laser (532 nm). Raman spectra of light scattering in dye-labelled cells on glass, in focal contacts attaching living dye-labelled cells to LB-MCNT-DTP-coating, and in PI molecules intercalated in the nanoheterostructures under investigation are shown in figs. 7, 8. The resonant Raman light scattering spectrum of PI molecules bound to the hydrophobic sites of the cells grown on glass has a maximum at frequency 2600 cm$^{-1}$ (see fig. 7, curve 0PI_C and fig. 8, curve 1).

Matching curves 0PI_C, 1_C, and 1PI_C in fig 7 one can see, that the parametric resonance of PI vibrational and plasmon modes on eigenfrequencies of free charge carriers in the LB-DTP-monolayes results in surface-enhanced resonant Raman scattering of PI. It follows from comparison of spectra 2 and 2PI in fig 7 in frequency range 1600-2300 cm$^{-1}$, the intercalation of PI molecules into MCNT-containing LB-layer does practically not intensify the luminescence of the sample. This is due to PI fluorescence quenching in complexes PI/MCNTs.

The cellular monolayer grown on the surface of LB-MCNT-DTP-film without PI molecules is characterized by elevated level of fluorescence after PI addition to cultural medium. Near the PI scattering maximum 2600 cm$^{-1}$ the scattering of the LB-MCNT-DTP-film with cells (spectrum 2PI_C in fig. 7) is more intensive than without cells (spectrum 2PI in fig. 7). Fluorescence of PI molecules intercalated into subcellular membrane structures gives an additional contribution into the scattering intensity.

According to fig. 8, the intensity of PI fluorescence in cells attached to the LB-MCNT-DTP-support are significantly enhanced at LB-DTP-monolayer plasmon frequencies, and, respectively, PI Raman spectrum situated at left from frequency 2300 cm$^{-1}$ increases sharply. Spectrum of PI in the case of glass support (fig. 8, curve 1) has a shoulder at frequency 3600 cm$^{-1}$.

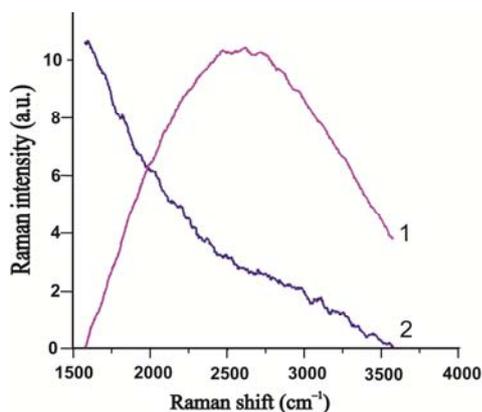

Fig. 8 Spectra of PI resonance Raman light scattering in cells on glass and in focal contacts attaching living cells to LB-MCNT-DTP-coating, curve 1 and 2 in fig. 6b, respectively. Excitation by green laser (532 nm) with power 0.06 mW

So, as it has been noticed previously, PI molecule fluorescence is quenched by water molecules. Then the maximal intensity of fluorescent probe emission takes place in the hydrophobic sites between cells and LB-MCNT-DTP film. These peculiarities of the confocal images of PI distribution on the LB-MCNT-coating will be used further to analyze structural organization of rat glioma cell monolayer.

*C. Visualization of Cellular Monolayer*

To visualize the structure of glioma cell monolayer grown on the studied LB-coating, cells were stained with propidium iodide. The resonance Raman scattering scanning of cell monolayer has been carried out at four frequencies: 1470, 1600, 2600, and 3600 cm$^{-1}$. The frequency 1470 cm$^{-1}$ is chosen to discover a proper luminescence of subcellular structures including focal contacts; the frequency 3600 cm$^{-1}$ – to exclude the MCNT Raman light scattering (spectral band 2D). The visualization frequency 1600 cm$^{-1}$ is chosen because the free charge carriers of metal-containing LB-DTP-film resonate parametrically at frequencies near left edge of the spectral PI band, according fig.7 and 8.

In figs. 9, 10 the typical images of PI distribution in cells grown on a cover glass (fig. 9a-c) and the LB-films (figs. 9d-i, 10) are shown. In control living cell monolayer on glass, the PI molecules form low-fluorescent complexes with hydrophobic focal contacts on cellular membrane (see image at 2600 cm$^{-1}$ in fig. 9b). Owing to weak cohesive interactions between cellular membrane and glass surface water molecules quench the vibrational mode of molecular groups everywhere except the edge region of cell touching to glass surface. PI luminescence is observed near cellular edge only indicating its tight surface adhesion. And simultaneously there exists a low Raman light scattering in some subcellular structures at scanning frequency 1470 cm$^{-1}$ as shown in fig. 9c.

As shown in fig. 9e, the LB-DTP-film without MCNTs enhances Raman light scattering of PI molecules intercalated both in focal adhesion sites and in the film. The last one blurs the image of adhered cell with prolate shape. More intensive luminescence and, correspondingly, more legible image of this cell in fig. 9g takes place at frequency 1470 cm$^{-1}$ because the spectral band of membrane luminescence at this frequency overlaps with plasmon mode and, consequently, the parametric resonance arises. In contrast to the glass, cell adhered to the LB-DTP-film touches to support surface tightly, as the surface-enhanced Raman scattering is observed in all areas where cell touches LB-DTP-film.

Let us investigate the morphology of rounded cell in fig. 9g. According to fig. 9h, the cellular body does not touch support everywhere except edges, because the focal contacts placed periphery are visualized in the focal plane of confocal microscope objective. The proper surface-enhanced scattering of subcellular structures at scanning frequency 1470 cm$^{-1}$ occurs in the same membrane sites that PI molecules scattering does, as one can see in fig. 9i. It is necessary to note that the focal contacts are circumferentially-spaced.

Uniform luminous background in fig. 9h at scanning frequency 2600 cm$^{-1}$ is likely to be the mode 2D of Raman light scattering in MCNTs.

Let study the morphology of extended adhered cell in fig. 10a at scanning frequencies 1600 and 3600 cm$^{-1}$. As shown in fig 10b at 1600 cm$^{-1}$, the well-defined high-intensity fluorescent image of cells with dark area, bright points of high-intensive PI luminescence, and highlight of near-surface membrane is registered by the surface plasmon enhanced resonance Raman scattering in subcellular structures with intercalated PI molecules. The dark areas lie outside of objective focal plane and are cellular nuclei regions curved by surface relief roughness (deep and shallow hollows in fig. 10b). Matching intensities of images in figs. 10b, c one can conclude that a MCNT-enhanced resonant Raman light scattering imaging is observed in fig. 10b.

The domains of high-intensive luminescence have diameters about 1 μm and are placed along periphery of cells (for all cells in scanning region) and around the dark areas. The distribution and sizes of found domains correspond to the focal adhesion sites of cells. Ordinary resonance Raman scattering by PI at 3600 cm$^{-1}$ in fig. 10c is not observed practically.

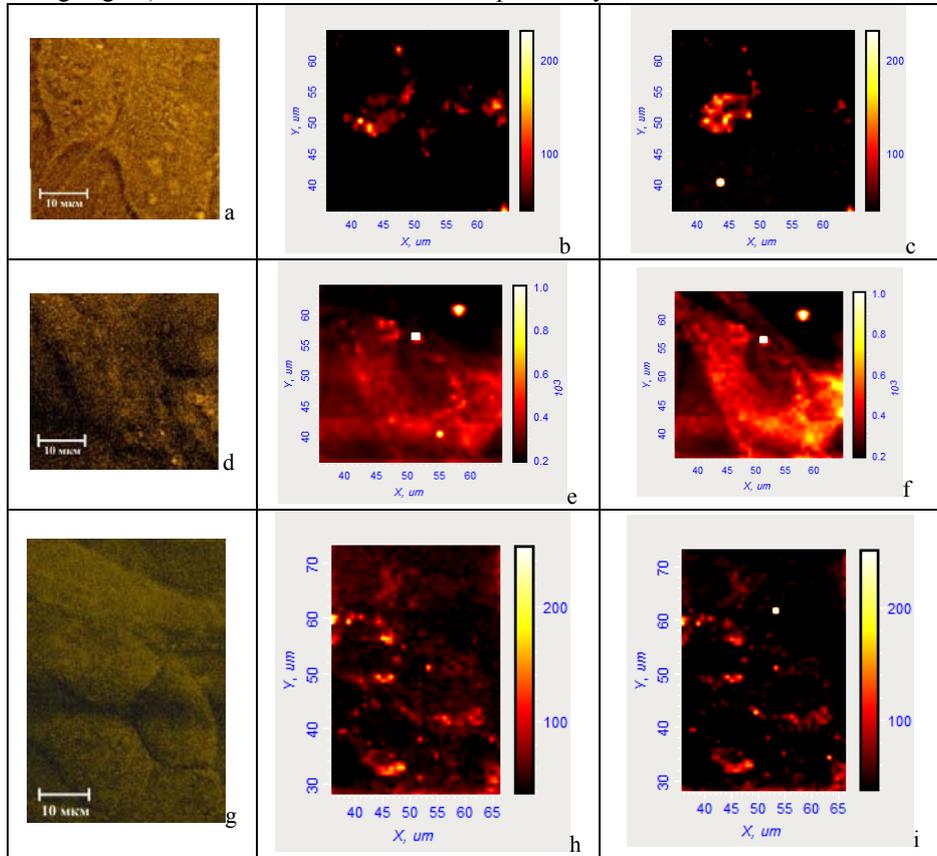

Fig. 9 Confocal microscope images of rat glioma cells grown on glass (positive control, 2 cells) (a–c), on LB-coating from five monolayers of DTP (d–f, 1 cell on film folds) and on LB-coating with MCNT (g–i, 4 cells). Excitation by green laser (532 nm), emission detection at frequencies (Raman shifts) $\upsilon$ = 2600 cm$^{-1}$ (in centre) and $\upsilon$ = 1470 cm$^{-1}$ (right). (In color)

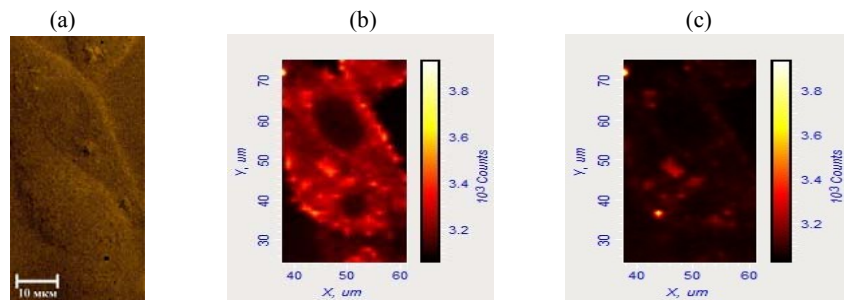

Fig. 10 Confocal microscope images of rat glioma cells grown on LB-coating with MCNT (2 cells). Excitation by green laser (532 nm), emission detection at frequencies (Raman shifts): 1600 cm$^{-1}$ (in centre) and 3600 cm$^{-1}$ (right). (In color)

## IV. CONCLUSION

To summarize, the existence of charge and energy transfer from ds-DNA to carbon nanotubes with a small number of walls has been found out. This charge transfer results in the appearance of free charge carriers transport in MCNT. The high-conducting MCNTs on metal-containing LB-DTP-support enhance the resonance Raman light scattering in propidium iodide fluorescent probe monolayer. The last phenomenon allows to image nanostructured surface of living cells grown on biocompatible coating such as thin metal- and MCNT-containing LB-film on nanoporous anodic alumina. Based on the results, the ability of DNA to bind to cellular membranes, and the ability of molecules PI to intercalate between pairs of nucleotide bases we can put forward the supposition about the participation of nucleotides in the mechanisms of focal adhesions of living cells to the extracellular matrix. It has been established that the diameter of focal adhesion sites is about 1 μm. The set of the integrin receptor zones is high-ordered and periodical one.

ACKNOWLEDGMENT

This study was supported by grants from the Ministry of Education and National Academy of Science, Republic of Belarus.